\documentclass{article}
\usepackage{amsfonts}
\usepackage{amsmath}

\input{tcilatex}
\begin{document}

\title{\textbf{Supersymmetric Dynamics of a Spin-1/2 Particle in an Extended
External Field}}
\author{G.S.Dias \\
IFES - Instituto Federal do Esp\'{\i}rito Santo - Campus Vit\'{o}ria - GFTA\\
gilmar@ifes.edu.br\\
}
\maketitle

\begin{abstract}
We consider a electron in a external field in $D=5$, through the Dirac
equation in the Galilean symmetry approach, and in the Lorentz symmetry
approach; from these we perform the nonrelativistic limit, then we procede
the supersymmetry of the same that is associated with the Galilean symmetry,
we identify as a supersymmetry sector from the quantum-mechanical dynamics,
and we got the algebra of fermionic charges. We naturally define as extra
electrical vector $E$, and interpret the terms of energy coming from the
fifth dimension. The energy from the fifth dimension, criate this extra
electrical vector $E$, associated with the fifth component of the external
electrical field $A$, this makes the energy flow from the fifth dimension to
the usual three-dimensional space, when some symmetries of the usual space
are broken, giving a preferential direction in the space, even though the
standard electrical and magnetic fields are null.
\end{abstract}

\section{\textbf{Introduction}}

\bigskip

Supersymmetry at low energies (non-relativistic limit) raises special
interest for a great deal of reasons, \cite{monopolosusyspinB}, \cite%
{aplicacaosusy}, \cite{capitulolivrosusy}, \cite{Gil2}, \cite%
{shapeinvariance}. If Supersymmetry is a symmetry of Nature, what we see
today must be its low-energy remnant through the breaking realised by means
of some mechanism \cite{susyvinculosquebra}, \cite{spinsusyB}, \cite%
{susyquebramecanicaquantica}. In this limit, the underlying field theory
should approach a Galilean invariant supersymmetric field theory and, by the
Bargmann superselection rule \cite{Bergmann}, such a field theory should be
equivalent to a supersymmetric Schr\"{o}dinger equation in each particle
number sector of the theory. One proposes a particle model based on the
Dirac equation in D=5, but we show some detail about dimensions less than 5.
This construction is embedded in a supersymmetric model consisting of an
extended Wess-Zumino model with 2 matter chiral superfields. Analyzing this
model in the low-energy limit, we proceed to the non-relativistic regime,
dropping out two degrees of freedom of the spinor that correspond to the
weak component. The system is identified as a sector of Galilean invariant
supersymmetric model. In some details this is equivalent work in the
approach of Galilean symmetry or Lorentzean; We can move from one approach
to another performing a linear transformation. In this way taking some
conditions, the supersymmetry model can cover both approachs. Here after
perform the supersymmetry, we show the interpretation of some parameters.

We can define an extra electrical field $E$, its origin is from the extra
dimension, it is generated from the extra component of the electric vector $%
A $, in the direction of the extra dimension, even though $A$ does not vary
in the direction of this extra dimension. This extra electrical field $E$
together with the presence of a charged particle with spin, for example an
electron, will obviously break the symmetry of space even though the
components of the electrical vector $A$, are zero in the usual space, ie not
in the extra-dimention. This extra electrical field $E$ in the presence of
spin and charge like the electron, generates two mechanisms by which energy
flows from the extra dimension to the usual space. The first mechanism is a
correction of the Pauli-Dirac Hamiltonian of the order of $A/m$ or $E/m^{2}$%
, where m is mass, we call this term $Psch$; it will give non-zero
contribution, if in the space there are a charged particle with spin-$1/2$
and a extra electrical vector $E$ not constant at the usual space. The
second mechanism that we call $Mtz$ will give a correction of order $A^{2}/m$
or $E^{2}/m^{3}$ since an electrical charge is in the presence of such extra
electrical field $E$ being constant or not in the usual space.

In this paper, we shall consider a particle of mass m and spin-$1/2$ in an
external field, \cite{monopolospinsusyA}, \cite{monopolosusyC}, \cite%
{ideiainicial},\cite{spinemsusy}. We build up a superspace action for this
model, read off the supersymmetry transformations of the component
coordinates and then obtain the supersymmetric charge operators \cite%
{calculocarga}. We analyse their algebra and pursue the investigation of the
possible existence of a central charge in the system. Then we perform the
interpretation of some results.

We set our discussion by first considering the case of (1+2) dimensions. The
Dirac action in such a space-time may be written as

\begin{equation*}
L=\overline{\Psi }(i\gamma ^{\mu }D_{\mu }-m)\Psi ,
\end{equation*}
where $\eta =(1,-1,-1)$ and $\gamma ^{0}=\sigma _{3}$, $\gamma ^{1}=i\sigma
_{2}$, $\gamma ^{3}=\sigma _{1}$, $\gamma _{3}=i\gamma _{0}\gamma _{1}$, $%
D_{\mu }\equiv\partial _{\mu }+ieA_{\mu }$. From the equation of motion,

\begin{equation*}
(i\gamma ^{\mu }D_{\mu }-m)\Psi =0,\text{ \ where \ \ }\Psi =\left( 
\begin{array}{c}
\psi _{1} \\ 
\psi _{2}%
\end{array}%
\right) ,
\end{equation*}

and with the approximation

\begin{equation*}
E+m-e\phi \cong 2m,\text{ \ \ and \ \ }\psi _{2}\cong -\frac{(D_{3}+iD_{1})}{%
2m}\psi _{1},
\end{equation*}

the Pauli Hamiltonian in 1+2 \cite{susydimensaodois} takes over the form:

\begin{equation}
H=\frac{1}{2m}\overrightarrow{\nabla }^{2}-\frac{ie}{2m}\overrightarrow{%
\nabla }.\overrightarrow{A}-\frac{ie}{m}\overrightarrow{A}.\overrightarrow{%
\nabla }-\frac{e^{2}}{2m}\left\vert \overrightarrow{A}\right\vert ^{2}-e\phi
.  \tag{1}
\end{equation}

The minimal coupling does not generate here any coupling term of the type
spin-magnetic field as it is the case for the Pauli Hamiltonian in D=1+3. We
can also show that this coupling term cannot appear from a non-minimal
coupling of the type:

\begin{equation*}
D_{\mu }=\partial _{\mu }+ieA_{\mu }+ig\widetilde{F}_{\mu },\text{ \ \ where
\ \ \ }\widetilde{F}_{i}=\frac{1}{2}\epsilon _{ijk}F^{jk}.
\end{equation*}%
If we call 1+2, (t,x,y) the usual space and add a extra dimention z to 1+2,
the dimension will be 1+3, (t,x,y,z), than we have in $H$ at less two new
terms related to corretions at Hamiltonian from the extra dimension, they
are:

\begin{equation}
Psch=\frac{ie}{2m}\sigma ^{3}\sigma ^{i}\overrightarrow{\nabla }^{i}(\Omega
),i=\{1,2\}\text{ \ \ \ \ \ \ \ \ \ \ \ \ and \ \ \ \ \ \ \ \ \ }Mtz=\frac{%
e^{2}}{2m}\left\vert \Omega \right\vert ^{2}  \label{Psch and Q D3}
\end{equation}%
where $\Omega $ is $A^{3}$, the extra component of the eletric vector A. In
the same way if we look at the dimention 1+1, (t,x) as our usual space, and
consider the direction z as an extra dimention,

\begin{equation}
Psch=\frac{ie}{2m}\sigma ^{3}\sigma ^{1}\overrightarrow{\nabla }^{1}(\Omega )%
\text{ \ \ \ \ \ \ \ \ \ \ \ \ and \ \ \ \ \ \ \ \ \ }Mtz=\frac{e^{2}}{2m}%
\left\vert \Omega \right\vert ^{2}  \label{Psch and Q D2}
\end{equation}

In (1+3)D, we see that the bosonic degree of freedom, $x^{i}(t)$ (position),
matches with the fermionic degree of freedom, $S^{i}(t)$ (spin). This makes
possible the construction of superfields and then of a manifestally
supersymmetric Lagrangian with linearly realised supersymmetry \cite%
{calculocarga}. However, in (1+2)D, the same does not occur; angular
momentum is not a vector any longer, the number of degreesof freedom of $%
x^{i}$ and the spin components no longer match and therefore we cannot in
this way build up superfields nor a supersymmetric Lagrangian in superspace.
We can try to work in (1+3)D, starting from Dirac's equation, taking the
non-relativistic limit and making a dimensional reduction to (1+2)D. But, we
see that it is not possible to drag into (1+2)D the analogue of the coupling
of spin-magnetic field that exists in (1+3)D \cite{susydimensaodois}. We
perform the supersymmetry in (1+4)D of this model starting from Dirac
Equation in the approach of Lorentz and Galilean Symmetry where we may
propose a match between space coordinates and the 5D spin, getting to an
extended eletromagnetic-like field.\bigskip

\section{Pauli-Dirac in the Lorentz Approach, L-A}

In the approach of Lorentz covariance we start from Dirac Lagrangean of a
electron in a extended electromagnetic field, and we obtain the Pauli-Dirac
Hamiltonian, the non-relativistic limit with minimal coupling.

We consider the following action in D=4+1,

\begin{equation}
L=\overline{\Psi }(i\Gamma \hat{^{\mu }}D_{\hat{\mu}}-m)\Psi ;
\label{Ldirac5d}
\end{equation}

where we define: 
\begin{equation}
\mu \in \{0,1,2,3,4\},\quad \eta =(+,-,-,-,-),\quad  \label{Definicao1}
\end{equation}

\begin{equation*}
x^{\mu }=(t,x,y,z,w),\quad F_{\mu \nu }=\partial _{\mu }A_{\nu }-\partial
_{\nu }A_{\mu },
\end{equation*}

\begin{equation*}
D_{\mu }=\partial _{\mu }+ieA_{\mu },\quad A_{\mu }=(A_{0},A_{i},\Lambda ),%
\text{ \ }A^{4}=\Omega =-\Lambda ,\text{ }\left\{ \Gamma ^{\mu },\Gamma
^{\nu }\right\} =2\eta ^{\mu \nu },
\end{equation*}

\begin{equation*}
\partial _{i}\Longleftrightarrow \overrightarrow{\nabla },\quad \partial
_{0}\Longleftrightarrow \frac{\partial }{\partial t},\quad
E^{i}\Longleftrightarrow \overrightarrow{E},\quad B^{i}\Longleftrightarrow 
\overrightarrow{B},\quad
\end{equation*}

\begin{equation*}
F_{04}=\mathcal{E},\quad F_{4i}=\mathcal{B}_{i},\quad F_{ij}=-\epsilon
_{ijk}B^{k},\quad F_{0i}=E_{i},
\end{equation*}

where $i,j,k\in \{1,2,3\}$. Notice that the scalar, $\mathcal{E}$, and the
vector, $\overrightarrow{\mathcal{B}}$, accompany the vectors $%
\overrightarrow{E}$ and $\overrightarrow{B}$, later on to be identified with
the electrical and magnetic fields, respectively. Our explicit
representation for the gamma-matrices is given in the Appendix 1. We set: 
\begin{equation*}
\Psi =\left( 
\begin{array}{c}
\psi _{1} \\ 
\psi _{2} \\ 
\psi _{3} \\ 
\psi _{4}%
\end{array}%
\right) ,\quad \overline{\Psi }=\Psi ^{\dagger }\Gamma ^{0}
\end{equation*}

We now take the equation of motion for ${\Psi }$ from the Euler-Lagrange
equations (\ref{Ldirac5d}). We suppose a stationary solution, 
\begin{equation*}
\Psi (x,t)=\exp (i\epsilon t)\chi (x);
\end{equation*}%
here, we are considering that the external field does not depend on t, and $%
A_{0}=0$ and that $\ \ \partial _{4}(A_{\mu })=0,$ then we obtain

\begin{equation}  \label{definicao2}
\overrightarrow{E}\Longleftrightarrow F_{0i}\equiv \partial _{0}A_{i}
-\partial _{i}A_{0}=0;
\end{equation}

\begin{equation*}
\mathcal{E}\equiv F_{04}\equiv \partial_{0} A_{4}-\partial_{4}A_{0}=0;
\end{equation*}

\begin{equation*}
F_{ij}=\partial _{i}A_{j}-\partial _{j}A_{i}\equiv \epsilon _{ijk}B_{k};
\end{equation*}

\begin{equation*}
F_{i4}=\partial _{i}A_{4}-\partial _{4}A_{i}\equiv \partial _{i}\Lambda
\equiv -\mathcal{B}_{i}.
\end{equation*}%
Then, considering the non-relativistic limit of the reduced theory, the
equations of motions read:

\begin{equation}
(\epsilon +m)I_{_{2}}\chi _{1}+(-i\sigma ^{i}(\partial
_{i}+ieA_{i})+iI_{_{2}}eA_{4})\chi _{2}=0.  \label{eqespinorial}
\end{equation}%
\begin{equation*}
(i\sigma ^{i}(\partial _{i}+ieA_{i})+iI_{_{2}}eA_{4})\chi _{1}+(-\epsilon
+m)I_{_{2}})\chi _{2}=0.
\end{equation*}%
therefore, we see that $\chi =\left( 
\begin{array}{c}
\chi _{1} \\ 
\chi _{2}%
\end{array}%
\right) $ loses two degrees of freedom, described by the "weak" spinor, $%
\chi _{2}$. So the Pauli-like Hamiltonian we get to reads as below:

\begin{equation}
H=\frac{1}{2m}\left[ -\left( \overrightarrow{\nabla }-ie\overrightarrow{A}%
\right) ^{2}-e\overrightarrow{\sigma }\left( \overrightarrow{\nabla }\times 
\overrightarrow{A}\right) \right.  \label{Hpauli54dn1}
\end{equation}

\begin{equation*}
\left. +e\overrightarrow{\sigma }\overrightarrow{\nabla }%
(A_{4})+e^{2}(A_{4})^{2}\right] ,
\end{equation*}
A similar term that relates energy from the extra dimention, like at (\ref%
{Psch and Q D2}) and (\ref{Psch and Q D3}) is present here, 
\begin{equation}
Psch=-\frac{e}{2m}\sigma ^{i}\overrightarrow{\nabla }^{i}(\Omega ),\text{ \
\ \ }i=\{1,2,3\}\text{ \ \ \ \ \ and \ \ \ \ \ \ }Mtz=\frac{e^{2}}{2m}%
\left\vert \Omega \right\vert ^{2}  \label{Psch and Q D4}
\end{equation}%
where we used (\ref{Definicao1}).

If we define, $p_{i}=-i\overrightarrow{\nabla }$ ($\hbar =1$) and $\left( 
\overrightarrow{\nabla }\times \overrightarrow{A}\right) ^{k}$ $=%
\overrightarrow{B}^{k}$ then the Hamiltonian (\ref{Hpauli54dn1}) will become

\begin{equation*}
H=\frac{1}{2m}\left[ \left( p_{i}-eA_{i}\right) ^{2}-eB^{i}\sigma ^{i}-e%
\mathcal{B}_{i}\sigma ^{i}+e^{2}\Lambda ^{2}\right] ;
\end{equation*}

\begin{equation}
H=\frac{\left( p_{i}-eA_{i}\right) ^{2}}{2m}+\frac{eB^{i}S_{i}}{m}-\frac{e%
\mathcal{B}_{i}S^{i}}{m}+\frac{e^{2}\Lambda ^{2}}{2m};\text{ onde }S^{i}=%
\frac{\sigma ^{i}}{2};  \label{PauliDirac LA}
\end{equation}

the corresponding Lagrangian is given by:

\begin{equation}
L=\frac{m}{2}(\dot{x}_{i})^{2}+\frac{i}{2}\psi _{i}\dot{\psi}_{i}+eA_{i}\dot{%
x}_{i}+  \label{Lpauli54dn1}
\end{equation}%
\begin{equation*}
\frac{ie}{2m}B_{i}\epsilon _{ijk}\psi _{j}\psi _{k}+\frac{ie}{2m}\mathcal{B}%
_{i}\epsilon _{ijk}\psi _{j}\psi _{k}-\frac{e^{2}\Lambda ^{2}}{2m},
\end{equation*}%
where the dot is a derivative with respect to t. We can also write

\begin{equation}
L=\frac{m}{2}(\dot{x}_{i})^{2}+\frac{i}{2}\psi _{i}\dot{\psi}_{i}+eA_{i}\dot{%
x}_{i}+  \label{Lpauli54dn1spin}
\end{equation}%
\begin{equation*}
-\frac{eB_{i}S_{i}}{m}-\frac{e\mathcal{B}_{i}S_{i}}{m}-\frac{e^{2}\Lambda
^{2}}{2m},
\end{equation*}%
where we identify the spin as the product below:

\begin{equation*}
S_{i}=-\frac{i}{2}\epsilon _{ijk}\psi _{j}\psi _{k}.
\end{equation*}%
The spin (angular momentum) is built up from the bosonic, but it is of
fermionic nature. In fact, we can check, with the help of the canonical
anti-comutation relations for the fermions (these shall be written down
later, in eq \ref{Relacaocomutacaon1}), that the algebra $%
[S_{i},S_{j}]=i\epsilon _{ijk}S_{k}$ is satisfied.

\section{Pauli-Dirac in the Galilean Approach, G-A}

Here we show a brief review of the formalism of Galilean covariance and the
Galilean Dirac Lagrangean of an electron in an extended electromagnetic
field, also we will show the Galilean Pauli-Dirac Hamiltonian, the
non-relativistic limit, with minimal and non-minimal coupling, when the spin
is considered. More details can be found in \cite{Montigny},\cite{GalDirac},%
\cite{GalileanCovariance}. We denote the coordinates in 4+1 dimensional flat
space-time $x^{\mu }=\left( x^{1},x^{2},x^{3},x^{4},x^{5}\right) $ and
choose the metric tensor by $\eta ^{\mu \nu }=diag(-1,-1,-1,1,-1)$, where $%
x^{4}=t_{r}$ \ is the usual time and $x^{5}$ is the extra dimension. We now
go to the light cone coordinates, were we denote the coordinates in 4+1
dimensional flat space-time $x^{\mu }=\left( x^{1},x^{2},x^{3},t,s\right) $,
by the linear transformation:%
\begin{equation}
t=\frac{i}{\sqrt{2}}(x^{4}-x^{5})\ \text{and }s=\frac{i}{\sqrt{2}}%
(x^{4}+x^{5})  \label{transform}
\end{equation}%
and the inversion \ $x^{4}=-\frac{i}{\sqrt{2}}(s+t)$ \ and $x^{5}=-\frac{i}{%
\sqrt{2}}(s-t);$ In the light-cone coordinates, the metric tensor has the
following form:

\begin{equation}
\left( g^{\mu \nu }\right) =\left( 
\begin{array}{ccccc}
1 & 0 & 0 & 0 & 0 \\ 
0 & 1 & 0 & 0 & 0 \\ 
0 & 0 & 1 & 0 & 0 \\ 
0 & 0 & 0 & 0 & -1 \\ 
0 & 0 & 0 & -1 & 0%
\end{array}%
\right) .  \label{Galilean Metric}
\end{equation}%
The set of all transformations that made $g^{\mu \nu }$ invariante is the
Galilean Group, they are: 
\begin{equation}
\begin{array}{rl}
\mathbf{x}^{\prime }= & R\mathbf{x}+\mathbf{v}t+\mathbf{a}, \\ 
t^{\prime }= & t+b, \\ 
s^{\prime }= & s+\left( R\mathbf{x}\right) \cdot \mathbf{v}+\frac{1}{2}%
\mathbf{v}^{2}t,%
\end{array}
\label{GalTrans}
\end{equation}%
where $R$ is a $3\times 3$ rotation matrix, $\mathbf{v}$ is the relative
velocity, $\mathbf{a}$ and $b$ are the space and time translations,
respectively. We note that the transformation in Eq. (\ref{GalTrans}) leaves
invariant the scalar product $g_{\mu \nu }dx^{\mu }dx^{\nu }.$ Now the
relativistic energy-momentum $p_{\mu }=\left( \mathbf{p},-\mathcal{E}%
,-m\right) $,\ the relationship for the massless particles $p^{\mu }p_{\mu
}=-2p_{s}p_{t}+\overrightarrow{p}^{2}=0$ can be rewritten if we write $%
p_{s}=-m$ and \ $p_{t}=-\mathcal{E},$ as

\begin{equation}
\mathcal{E}=\frac{1}{2m}\overrightarrow{p}^{2}  \tag{4}
\end{equation}%
which is nothing but the expression of the kinetic energy for the
non-relativistic particle with mass $m$ in 4 dimensional space, 5 space-time
dimensions. This is the natural interpretation of variables $t$ and $s,$
what is not more than the mix of the dimensions $x^{4}=t_{r}$ usual time and 
$x^{5};$ that interpretation follow from the canonically conjugate
variables: The five-momentum $p_{\mu }$.

The Galilean Dirac equation 
\begin{equation*}
\gamma ^{\mu }\partial _{\mu }\Psi =0,\;\;\;\;\;\overline{\Psi }\gamma ^{\mu
}\overset{\leftarrow }{\partial }_{\mu }=0,
\end{equation*}%
where $\gamma ^{\mu }\partial _{\mu }=\mathbf{\gamma }\cdot \nabla +\gamma
^{t}\partial _{t}+\gamma ^{s}\partial _{s}$ and the basic field is the
spinor, $\Psi =\left( 
\begin{array}{c}
\vartheta _{+} \\ 
\vartheta _{-}%
\end{array}%
\right) ,$ other notations\ about that section can be seen in appendix 1.

\QTR{uline}{The Galilean-Dirac Lagrangian: 
\begin{equation*}
\mathit{L}=\frac{1}{2}\overline{\Psi }\left( \gamma ^{\mu }\partial _{\mu
}\right) \Psi -\frac{1}{2}\overline{\Psi }\left( \gamma ^{\mu }\overset{%
\leftarrow }{\partial }_{\mu }\right) \Psi .
\end{equation*}%
}

The Galilean-Dirac Lagrangian for a particle in a electromagnetic field can
be obtained directly with the following definitions, \QTR{uline}{the minimal
coupling $p_{\mu }\rightarrow \pi _{\mu }=p_{\mu }-eA_{\mu }$, together
another non-minimal coupling, ${\pi _{i}}\rightarrow {\pi _{i}}+C_{i}\eta $,
where $\mathbf{C}=\frac{ie}{4m}\mathbf{E}$. This leads to }$\mathbf{p}%
\rightarrow \mathbf{p}-e\mathbf{A}+ie\frac{\mathbf{E}}{4m}\eta ,$ $%
p_{4}\rightarrow p_{4}-eA_{4},\ \ p_{5}\rightarrow p_{5}-eA_{5},$ 
\QTR{uline}{\ where we use the $5$-potential $A_{\mu }=\left(
A_{i},A_{4},A_{5}\right) =\left( \mathbf{A},-\phi _{m},-\phi _{e}\right) $.
The Galilean }Pauli-Dirac Hamiltonian, the non-relativistic limit, with
minimal and non-minimal coupling can be easily produced, this result is in 
\cite{Montigny}.

Using here in G-A the same conditions used for the external field $A$ in
L-A, like $A^{0}=0$ in L-A will be $\phi _{m}=-\phi _{e}$ in G-A by
transformations the (\ref{transform}). Also the electrical field we fix, 
\QTR{uline}{$\mathbf{E=0,}$ giving us} from \cite{Montigny} the following
Hamiltonian to both, \QTR{uline}{large and small components of $\Psi ,$}%
\begin{equation}
\mathcal{E}\vartheta _{\pm }=\left( \frac{\left( \mathbf{p}-e\mathbf{A}%
\right) ^{2}}{2m}+\frac{e}{m}\mathbf{S}\cdot \mathbf{B}\right) \vartheta
_{\pm }.  \label{PauliDirac GA}
\end{equation}%
That Hamiltonian is a sector that one in L-A, eq (\ref{PauliDirac LA}). We
could imagine that in G-A it is not possible to represent the interactions
from the fifth dimension in this way like we represente in G-A at eq (\ref%
{PauliDirac LA}), because the terms $Psch$ and $Mtz$ eq (\ref{Psch and Q D4}%
) that is responsable for this do not appear here, ie we have no
interactions from the fifth dimension, but in the last section we will see
that they can be generated with the inclusion of a dimensional factor in the
non-minimal coupling even though the usual vector field vanishes, 
\QTR{uline}{$\mathbf{E=0,}$because we will be able to define any extra
electrical field }$\QTR{uline}{E}$\QTR{uline}{\ generated by the energy from
the extra-dimension}.

If the usual electrical field $E$ is not zero, we know from \cite{Montigny}
that the Hamiltonian (\ref{PauliDirac GA}) in some conditions (like static
field and others) has the following two terms of correction:%
\begin{equation}
MR\mathcal{=}\frac{e^{2}}{32m^{3}}\mathbf{E}^{2}\text{\ \ and\ \ \ \ }OR=%
\frac{e}{8m^{2}}\nabla \cdot \mathbf{E}  \label{Correcoes PauliDirac}
\end{equation}

\section{N=1 Galilean Supersymmetry}

To render our discussion more systematic, we think it is advisable to set up
a superfield approach. We can define the N=1-supersymmetric model in analogy
with the model presented above, eq.(\ref{Lpauli54dn1spin}). We start
defining the superfelds by

\begin{equation}
\Phi _{i}(t,\theta )=x_{i}(t)+i\theta \psi _{i}(t)\quad \Sigma (t,\theta
)=\xi (t)+\theta R(t),  \label{Definicao3}
\end{equation}

\begin{equation*}
\Lambda (x)=A_{4}(x) \quad \Lambda (\Phi _{s})=\Lambda (x) +i(\partial
_{j}\Lambda (x))\theta \psi _{j}.
\end{equation*}

The supercharge operators and the covariant derivatives are given by: 
\begin{equation}
Q=\partial _{\theta }+i\theta \partial _{t}\quad D=\partial _{\theta
}-i\theta \partial _{t}\quad H=i\partial _{t}.  \label{Operadoresn1}
\end{equation}

Then, the N=1-supersymmetric Lagrangian that generates the Galilean
Hamiltonian (\ref{PauliDirac GA}) can be written in superfields as: 
\begin{equation}
\mathcal{L}=\frac{i}{2}\dot{\Phi}_{i}D\Phi _{i}+ie(D\Phi _{i})A_{i}(\Phi )+%
\frac{1}{2}\Sigma D\Sigma +  \label{Lsusyn1}
\end{equation}%
\begin{equation*}
-e\Sigma \Lambda (\Phi )+\frac{ie}{2}\epsilon _{ijk}\partial _{i}\Lambda
(\Phi )\Phi _{j}D\Phi _{k},
\end{equation*}%
where we have set $m=1$.

The commutators and anticommutators for the superfield components are

\begin{equation}  \label{Relacaocomutacaon1}
\left[ \psi _{i,}\psi _{j}\right] _{+}= \delta _{ij},\quad \left[ \xi , \xi %
\right] _{+}=-2i,\quad \left[ x_{i,}p_{j}\right] =i\delta _{ij},
\end{equation}

\begin{equation*}
\left[ \psi _{i,}\xi \right] _{+}=0;
\end{equation*}
the other commutators vanish.

The component field R does not have dynamics; then, we use the equation of
motion to remove it from the Lagrangian: 
\begin{equation}
R=e\Lambda .  \label{Eqr}
\end{equation}

The supersymmetric action is then 
\begin{equation*}
\mathcal{S}=\int dtd\theta \mathcal{L}
\end{equation*}

We have written the supersymmetric Lagrangian in components, \cite%
{calculocarga}, under the form,

\begin{equation*}
\mathcal{L}=K+\theta L
\end{equation*}
where we define $K$ and $L$ as follows:

\begin{equation*}
K=-\frac{1}{2}\dot{x}_{i}\psi _{i}-e\psi _{i}A^{i}+\frac{\xi R}{2}-e\xi
\Lambda -\frac{e}{2}\epsilon _{ijk}\mathcal{B}_{i}x_{j}\psi _{k}
\end{equation*}

\begin{equation}  \label{Lsusycomponentesn1}
L=\frac{1}{2}\dot{x}_{i}\dot{x}_{i}- \frac{1}{2}i\dot{\psi }_{i}\psi _{i} +e%
\dot{x}_{i}A_{i}-\frac{ie}{2}F_{ij}\psi _{i}\psi _{j}+
\end{equation}

\begin{equation*}
+\frac{i}{2}e\xi \dot{\xi }+\frac{R^{2}}{2}-eR\Lambda +ie\xi \psi _{j}%
\mathcal{B}_{j}+
\end{equation*}
\begin{equation*}
+\frac{e}{2}\epsilon _{ijk}\mathcal{B}_{i}x_{j}\dot{x}_{k} -\frac{ie}{2}%
\epsilon _{ijk}\mathcal{B}_{i}\psi _{j}\psi _{k} -\frac{ie}{2}\epsilon
_{ijk}\psi _{r}\partial _{i} \mathcal{B}_{r}x_{j}\psi _{k}
\end{equation*}

Also from $L$ in (\ref{Lsusycomponentesn1}), we can read the Hamiltonian:

\begin{equation}
H=\frac{1}{2}(p^{i}-eA^{i})^{2}+\frac{ie}{2}\epsilon _{ijk}B_{k}\psi
_{i}\psi _{j}+  \label{Hsusycomponentesn1}
\end{equation}

\begin{equation*}
-\frac{ie}{2}\epsilon _{ijk}\mathcal{B}_{i}\psi _{j}\psi _{k} -\frac{ie}{2}%
\epsilon _{rji}(\partial _{r}\mathcal{B}_{k})x_{j} \psi _{k}\psi _{i}+ie 
\mathcal{B}_{i}\xi\psi _{i}+
\end{equation*}

\begin{equation*}
-\frac{e}{2}\epsilon _{jki}\mathcal{B}_{j}x_{k}(p_{i}-eA^{i}) +\frac{e^{2}}{8%
}\epsilon _{jki}\epsilon _{rni}\mathcal{B}_{j}\mathcal{B}_{r}x_{k}x_{n}+%
\frac{e^{2}\Lambda ^{2}}{2};
\end{equation*}
in this equation, we have eliminated the component-field without dynamical
character.

\subsection{The supersymmetric charge}

Acting with the supercharge operator (\ref{Operadoresn1}) on the superfields
(\ref{Definicao3}), we can obtain the supersymmetry transformations of the
components of the fields; they are:

\begin{equation}  \label{Transfn1}
\delta \psi ^{i}=\epsilon \dot{x}^{i},\quad \delta \xi =-i\epsilon R,\quad
\delta R=-\epsilon \dot{\xi },\quad \delta x^{i}=-i\epsilon \psi ^{i}.
\end{equation}

From the supersymmetric transformations and the Lagrangian (\ref%
{Lsusycomponentesn1}), we can analytically calculate the supercharge,
through the Noether's theorem. The charge operator comes out to be

\begin{equation}  \label{Cargaanalitican1}
Q=\psi _{i}\dot{x}_{i}+\frac{1}{2}(1-i)\xi R- e\xi \Lambda.
\end{equation}

The supercharge algebra reads

\begin{equation*}
\left[ Q,Q\right] _{+}=2H
\end{equation*}%
where H is the Hamiltonian of eq. (\ref{Hsusycomponentesn1}). The
quantization was already performed, \cite{quantizacaosusyvinculo}.

This supersymmetry includes both approaches of this model, G-A and L-A, the
terms $Psch$ and $Mtz$ that make the correction to the Hamiltonian in L-A,
was decribed in the supersymmetric model. It is possible to give a
interpretation about the physical mechanism by which the corretion terms $%
Psch$ and $Mtz$ of the Hamiltonian (\ref{PauliDirac LA}), allow the energy
flow from the fifth dimension (extra dimension) to the usual space. And from
this new interpretation, we generate this two corretion terms $Psch$ and $%
Mtz $ in the G-A.\bigskip

\section{Extra Electrical Field from the Extra Dimension}

The Hamiltonian eq (\ref{Hpauli54dn1}) has two correction factors from the
extra dimension, fifth dimension, which appear also in the Hamiltonian from
the supersymmetry eq (\ref{Hsusycomponentesn1}), $Psch$ term with order of $%
\frac{A_{4}}{m}$; and the $Mtz$ term with order $\frac{A_{4}^{2}}{m}$. Both
terms also appear in the Hamiltonian in lower dimensions, which have an
extra dimension, as in 1+2 or 1+3, eq (\ref{Psch and Q D2}), (\ref{Psch and
Q D3}) . The Energy from the fifth dimension only flows to the usual
three-dimensional space by the mechanism $Psch$, if the symmetry of the
usual space is broken into three simultaneous aspects. It is due to the
external electrical vector component$\ (A)$ on the extra dimension is not
homogeneous on the usual space (not constant ), generating a preferred
direction on the usual space. And the other two factors are due the presence
of electrically charged particle with spin 
$\frac12$%
. By the mechanism $Mtz$, the symmetry of the usual space is broken allowing
the energy flow from the extra dimension to the usual space, since exist an
electrically charged particle, in presence of a extra electrical field $%
\overrightarrow{E}$ in 3 dimension, or in presence of a pseudo-electrical
field $\overrightarrow{E}$ in 2 or 1 dimention. Both the extra electrical
field and the pseudo-electrical field $\overrightarrow{E}$ are generated
from the extra-dimension by the existence of a nonzero component of the
eletrical vector $A$ in the extra dimension, being $A$ breaking the symmetry
of space or not. \ Based on the corrections on the Hamiltonian in the G-A eq
(\ref{PauliDirac GA}), in terms of a usual electrical field $E$, eq (\ref%
{Correcoes PauliDirac}), we use exactly the same term, which will differ
only by a multiplicative constant related to the degree of freedom of the
particle, ie the dimension of usual space. In this sense, at the case of 1
+2 dimension, the correction terms of the Hamiltonian in terms of the
electrical field is the same given by, $MR$ and $OR$ \ to the usual
electrical fields,%
\begin{equation}
MR\mathcal{=}\frac{e^{2}}{32m^{3}}\mathbf{E}^{2}\text{\ \ and\ \ \ \ }OR=%
\frac{e}{8m^{2}}\nabla \cdot \mathbf{E}
\end{equation}%
and the definition of the extra electrical field, which is actually a pseudo
vector, it is invariant by parity, is 
\begin{equation}
E^{1}=i4m\Omega \sigma ^{3}\sigma ^{1}
\end{equation}%
that creates the terms $Psch=\frac{ie}{2m}\sigma ^{3}\sigma ^{1}%
\overrightarrow{\nabla }^{1}(\Omega )$ \ and $Mtz=\frac{e^{2}}{2m}\left\vert
\Omega \right\vert ^{2}$, the proper correction of the Hamiltonian eq (\ref%
{PauliDirac GA}) in G-A, due to the influence of the energy of the extra
dimension. Although the usual electrical field is zero, the extra pseudo
electrical field generated by the energy from the extra dimension could rise.

In a general way in 1+2 and 1+3 dimension we can define the pseudo electical
field by, 
\begin{equation}
E^{i}=\frac{i4m}{\sqrt{d}}\Omega \sigma ^{3}\sigma ^{i}\text{ },\text{\ }%
i=\{1,2\},d=2\text{ to 1+3 or \ }i=\{1\},d=1\text{ to 1+2}
\label{pseudo vetor eletrico}
\end{equation}%
The correction terms of the Hamiltonian on the dimension 1+2, 1+3 and 1+4 in
terms of the extra electrical field is similar to $MR$ and $OR$ 
\begin{equation}
MR\mathcal{=}\frac{e^{2}}{32m^{3}}\mathbf{E}^{2}\text{\ \ and\ \ \ \ }OR=%
\sqrt{d}\frac{e}{8m^{2}}\nabla \cdot \mathbf{E}  \label{correcao energia}
\end{equation}%
\ on 1+3 dimension we have the correction terms $Psch=\frac{ie}{2m}\sigma
^{3}\sigma ^{i}\overrightarrow{\nabla }^{i}(\Omega )$ \ and $Mtz=\frac{e^{2}%
}{2m}\left\vert \Omega \right\vert ^{2}$ . where $d$ is the dimensional
degree of freedom of the particle, the usual space dimension were the
particles actually move. In the 1+4 dimention the extra electrical field is
really a vector defined by,%
\begin{equation}
E^{i}=-\frac{4m}{\sqrt{3}}\Omega \sigma ^{i}\text{ },\text{\ }i=\{1,2,3\}
\end{equation}%
that generates the terms $Psch=-\frac{e}{2m}\sigma ^{i}\overrightarrow{%
\nabla }^{i}(\Omega )$ $\ \ i=\{1,2,3\}$ and $Mtz=\frac{e^{2}}{2m}\left\vert
\Omega \right\vert ^{2}$ the proper correction of the Hamiltonian eq (\ref%
{PauliDirac GA}) in G-A, due to the influence of the energy from the extra
fifth dimension. With this correction and conditions the Pauli-Dirac
Hamiltonian in G-A is the same one in L-A eq (\ref{Hpauli54dn1}), that is a
supersymmetric sector of eq (\ref{Hsusycomponentesn1}). So even though the
usual electrical field $E$ is zero in the usual three-dimensional space, the
existence of a nonzero extra electrical field will allow energy to flow from
the fifth dimension to the usual three-dimensional space, by the mechanism $%
Psch$ and $Mtz$.

\subsection{Conclusion}

We showed in two approaches, L-A and G-A the Pauli-Dirac Hamiltonian in 4+1
dimensions, and we performed the supersymmetry of the model inside some
conditions. also we could show that the term $Psch$ and the term $Mtz$
arises in a similar way in 1 +2, 1 +3 and 1 +4 dimension, but the extra
electrical field associated with these terms in 1 +2 dimensions and 1 +3 are
pseudo vectors, only in 1+3 dimension, it will be a vector. The reason for
this, is the same by which it was not possible to describe in a 1+2 the
supersymmetry here, using the positions of the bidimensional space as
bosonic variables and angular momentum, spin as fermionic variables. The
angular momentum in 1+2 dimension is not a vector it is a pseudovector, it
is invariant by parity, or equivalently one can say that a supersymmetry in
this way is not possible here, as was discussed previously by treating the 2
+1 dimension. The extra electrical field $E$ produces the correct terms in
eq (\ref{Psch and Q D4}) from the terms $MR$ and $OR$. The extra electrical
field $E$ from the extra dimension was defined in a natural way, leading to
a new result, which was not established for the Hamiltonian in G-A. The
factor $d$ that appears in the definition of the electrical vector and in
the pseudoelectrical vector reveals the dimension where the particle is
trapped, it is the usual dimension of space. In graphene the electron for
example is apparently confined to a space of dimension two, in this way the
third spatial coordinate would be extra. An experiment may show whether this
parameter $d$ has in fact physical means, in the case of graphene if
actually the electron is confined in a surface, d = 2. The existence of an
extra vector or extra pseudo-vector type electric is a simple way to see
that the space could contain more energy than usually we realize, it would
be hidden in other dimensions, and arise when the symmetry of space is
broken by some mechanism like $Psch$ or $Mtz.$

\section*{Appendix 1}

We quote below the $\Gamma $-matrices representing the Clifford algebra of
(1+4)D: 
\begin{equation*}
\Gamma ^{4}=i\Gamma ^{5}=\left( 
\begin{array}{cc}
0 & i_{2} \\ 
i_{2} & 0%
\end{array}%
\right) ;\quad \Gamma ^{i}=\left( 
\begin{array}{cc}
0 & \sigma ^{i} \\ 
-\sigma ^{i} & 0%
\end{array}%
\right) ;\quad \Gamma ^{0}=\left( 
\begin{array}{cc}
1_{2} & 0 \\ 
0 & -1_{2}%
\end{array}%
\right) .
\end{equation*}

\section*{Appendix}

Its adjoint \QTR{uline}{{is} given by $\overline{\Psi }=\Psi ^{\dagger
}\gamma ^{t_{r}}=\Psi ^{\dagger }\eta _{(+)}$, where $\vartheta _{+}$ and $%
\vartheta _{-}$ are the large and small components respectively, $\eta
_{(+)}=\gamma ^{t_{r}}=$}$\QTR{uline}{\frac{-i}{\sqrt{2}}\left( \gamma
^{s}+\gamma ^{t}\right) }$\QTR{uline}{, $\ \eta _{(-)}=\gamma ^{x^{4}}=%
\QTR{uline}{\frac{-i}{\sqrt{2}}\left( \gamma ^{s}-\gamma ^{t}\right) }$ and $%
\gamma ^{\mu }$, $\mu =1,2,3,t,s,$ are $4\times 4$ matrices that satisfy the
Clifford algebra }$\gamma ^{\mu }\gamma ^{\nu }+\gamma ^{\nu }\gamma ^{\mu
}=2g^{\mu \nu }$\QTR{uline}{. The representation here for these matrices, is
given by 
\begin{equation*}
\gamma ^{i}=\left( 
\begin{array}{cc}
\sigma ^{i} & 0 \\ 
0 & -\sigma ^{i}%
\end{array}%
\right) ,\qquad \gamma ^{t}=\sqrt{2}\left( 
\begin{array}{cc}
0 & 0 \\ 
-\mathbf{1} & 0%
\end{array}%
\right) ,\qquad \gamma ^{s}=\sqrt{2}\left( 
\begin{array}{cc}
0 & \mathbf{1} \\ 
0 & 0%
\end{array}%
\right) ,
\end{equation*}%
where $\sigma ^{i}$, $i=1,2,3$, are the Pauli matrices. }

\section*{ACKNOWLEDGEMENTS}

The author is grateful to Jos\'{e} Abdalla Helayel-Neto for pleasant and
clarifying discussions in supersymmetry. The author express his gratitude to
IFES, Instituto Federal do Espirito Santo, and the Sub-Reitoria of Research
at the IFES for the financial support, too the University of Alberta -
Department of Physics to provide a office. The author is grateful to Faqir C
Khanna and Marc de Montigny for pleasant and clarifying discussions in
algebra contraction and by the interesting proposals and suggestions who are
currently being generated and emerged from this work, immediately after the
first time that this work was presented to them, mainly in galilean
symmetry; and grateful to his friend Rick Dumbar to suggest corrections in
his english in this work.

\end{document}